\documentclass[a4paper, amsfonts, amssymb, amsmath, reprint, prl, aps,
         showkeys, nofootinbib, superscriptaddress, twoside, nosort, twocolumn]{revtex4-2}

\usepackage{lipsum}
\usepackage{float}
\usepackage{tabularx}
\usepackage{upgreek}
\usepackage{mathtools}
\usepackage{physics}
\usepackage{xcolor}
\usepackage{graphicx}
\usepackage[left=23mm,right=13mm,top=35mm,columnsep=15pt]{geometry}
\usepackage{adjustbox}
\usepackage{placeins}
\usepackage[T1]{fontenc}
\usepackage[english]{babel}
\usepackage[utf8]{inputenc}
\usepackage{csquotes}
\usepackage[mathscr]{eucal} 
\usepackage[globalcitecopy]{bibunits} 
\usepackage[pdftex, pdftitle={Article}, pdfauthor={Author}]{hyperref} 
\usepackage[capitalize]{cleveref}
\usepackage{multirow}
\usepackage{siunitx}
\usepackage{natbib}

\newcommand{\Sfive}{\mathrm{S\text{-}5}}
\newcommand{\Sseven}{\mathrm{S\text{-}7}}
\newcommand{\DeltaAl}{\Delta_{\mathrm{Al}}}
\newcommand{\DeltaBulk}{\Delta_{\mathrm{Bulk}}}
\newcommand{\DeltaSfive}{\Delta_{\mathrm{S\text{-}5}}}
\newcommand{\DeltaSseven}{\Delta_{\mathrm{S\text{-}7}}}
\newcommand{\fq}{f_{\mathrm{q}}}
\newcommand{\nth}{k}
\newcommand{\gammakept}{\gamma_{\mathrm{kept}}}
\newcommand{\gammarej}{\gamma_{\mathrm{rej}}}
\newcommand{\RO}{\mathrm{RO}}
\newcommand{\fRO}{f_{\mathrm{RO}}}
\newcommand{\tx}{t_{X}}
\newcommand{\twait}{t_{\mathrm{wait}}}
\newcommand{\tro}{t_{\mathrm{RO}}}
\newcommand{\tidle}{t_{\mathrm{idle}}}
\newcommand{\tcycle}{t_{\mathrm{cycle}}}
\newcommand{\trec}{t_{\mathrm{rec}}}
\newcommand{\Tone}{T_\mathrm{1}}
\newcommand{\Ttwoecho}{T_2^{\mathrm{E}}}
\newcommand{\TtwoRamsey}{T_2^{*}}

\newcommand{\ueV}{\mu\mathrm{eV}}
\newcommand{\s}{\mathrm{s}}
\newcommand{\ns}{\mathrm{ns}}
\newcommand{\us}{\upmu\mathrm{s}}
\newcommand{\ms}{\mathrm{ms}}
\newcommand{\nm}{\mathrm{nm}}
\newcommand{\um}{\upmu\mathrm{m}}
\newcommand{\mm}{\mathrm{mm}}
\newcommand{\cm}{\mathrm{cm}}

\newcommand{\minutes}{\mathrm{min}}
\newcommand{\hours}{\mathrm{h}}

\newcommand{\MHz}{\mathrm{MHz}}
\newcommand{\GHz}{\mathrm{GHz}}
\newcommand{\K}{\mathrm{K}}

\begin{document}
\title{
Qubit error bursts in superconducting quantum processors of Quantum Inspire: quasiparticle pumping and anomalous time dependence
}
\author{G.~R.~Di~Carlo}
\thanks{These authors contributed equally to this work.}
\affiliation{QuTech and Kavli Institute of Nanoscience, Delft University of Technology, P.O. Box 5046, 2600 GA Delft, The Netherlands}

\author{M.~Samiotis}
\thanks{These authors contributed equally to this work.}
\affiliation{QuTech and Kavli Institute of Nanoscience, Delft University of Technology, P.O. Box 5046, 2600 GA Delft, The Netherlands}

\author{A.~Kamlapure}
\affiliation{QuTech and Kavli Institute of Nanoscience, Delft University of Technology, P.O. Box 5046, 2600 GA Delft, The Netherlands}

\author{M.~Finkel}
\affiliation{QuTech and Kavli Institute of Nanoscience, Delft University of Technology, P.O. Box 5046, 2600 GA Delft, The Netherlands}

\author{N.~Muthusubramanian}
\affiliation{QuTech and Kavli Institute of Nanoscience, Delft University of Technology, P.O. Box 5046, 2600 GA Delft, The Netherlands}
\affiliation{Present address: AWS Center for Quantum Computing, 290 South Holliston Avenue, California Institute of Technology, Pasadena, CA 91106 USA}

\author{M.~W.~Beekman}
\affiliation{QuTech and Kavli Institute of Nanoscience, Delft University of Technology, P.O. Box 5046, 2600 GA Delft, The Netherlands}
\affiliation{Netherlands Organisation for Applied Scientific Research (TNO), P.O. Box 96864, 2509 JG The Hague, The Netherlands}

\author{N.~Haider}
\affiliation{QuTech and Faculty of Electrical Engineering, Mathematics, and Computer Science, Delft University of Technology, Mekelweg 4, 2628 CD Delft, The Netherlands}

\author{M.~S.~Moreira}
\affiliation{QuTech and Kavli Institute of Nanoscience, Delft University of Technology, P.O. Box 5046, 2600 GA Delft, The Netherlands}
\affiliation{Present address: Department of Electrical Engineering and Computer Science, Massachusetts Institute of Technology, Cambridge, MA 02139, USA}

\author{J.~F.~Marques}
\affiliation{QuTech and Kavli Institute of Nanoscience, Delft University of Technology, P.O. Box 5046, 2600 GA Delft, The Netherlands}
\affiliation{Present address: AWS Center for Quantum Computing, 290 South Holliston Avenue, California Institute of Technology, Pasadena, CA 91106 USA}

\author{B.~Segers}
\affiliation{QuTech and Kavli Institute of Nanoscience, Delft University of Technology, P.O. Box 5046, 2600 GA Delft, The Netherlands}
\affiliation{Present address: Quantware B.V., Elektronicaweg 10, 2628 XG Delft, The Netherlands}

\author{S.~Vallés-Sanclemente}
\affiliation{QuTech and Kavli Institute of Nanoscience, Delft University of Technology, P.O. Box 5046, 2600 GA Delft, The Netherlands}

\author{L.~DiCarlo}
\email{Corresponding author: l.dicarlo@tudelft.nl}
\affiliation{QuTech and Kavli Institute of Nanoscience, Delft University of Technology, P.O. Box 5046, 2600 GA Delft, The Netherlands}

\date{\today}
\begin{bibunit}[apsrev4-2]

\begin{abstract}

We investigate qubit error bursts in 5- and 7-transmon processors of similar design, fabrication and packaging, but with different types of qubit Josephson junctions. Measurements for each are performed in two refrigerators to discern device-specific from refrigerator-dependent characteristics. The duration and rate of bursts are device specific but within the range of prior experiments and consistent with ionizing radiation. We observe two unforeseen signatures specifically in the processor with Dolan junctions. First, increasing the rate of $\pi$ pulsing in the detection scheme shortens the recovery time to equilibrium, which is explained by a quasiparticle pumping mechanism. The second signature is an anomalous time dependence in the burst rate: a surge happens days or weeks after cooldown, followed by a strong suppression that persists until thermal cycling.

\end{abstract}
\maketitle

It is counterintuitive that high-energy radiation of cosmogenic origin and from local radioactivity, with as much as MeV-GeV per particle/photon, can cause the majority of qubits in a superconducting quantum processor to lose energy~\cite{McEwen22, McEwen24, Harrington24, Li24}. Yet, the various energy inter-conversion processes underlying this catastrophic effect are understood~\cite{Martinis21, Kaplan76}. Incident radiation sheds up to MeV in the form of phonons and substrate electron-hole pairs, with most of the latter quickly recombining to produce more phonons. Carrying $\sim90\%$ of the shed energy, phonons proliferate by frequency down-conversion and propagate across the substrate, partially transmitting into superconducting layers where they break Cooper pairs into quasiparticles (QPs). Through their continued interactions, phonons and QPs down-convert in a process culminating with $\sim60\%$ of the initial phonon energy as QPs with energy close to the superconducting gap $\Delta$ and the rest as phonons with energy $<2\Delta$. QPs on the electrodes of qubit Josephson junctions (JJs) can tunnel across by absorbing the qubit transition energy $\fq$~\cite{Martinis09, Catelani11}. Such qubit relaxation errors persist until the QP density near qubit JJs subsides through recombination and trapping mechanisms. However, $2\Delta$ phonons produced by recombination of QPs can re-break Cooper pairs, increasing the recovery time. Ultimately, most of the energy imparted by the radiation escapes the processor as $\leq2\Delta$ phonons via interconnects (AlSi wirebonds, In bumps, etc.) and clamping points.

The increased propensity for qubits to relax, made global by phonon propagation, is a source of correlated errors and consequently problematic for quantum error correction (QEC) and eventually fault-tolerant quantum computing. In fact, such error bursts already set the performance limit of state-of-the-art QEC experiments~\cite{Google21, Google25}. Several strategies are being pursued that can help mitigate the impact. Recent large-scale efforts involve moving experimental setups to deep underground facilities to reduce the flux of ionizing radiation of cosmogenic origin~\cite{Cardani21, Gusenkova22}. Device-level innovations include phonon traps~\cite{Patel17, Henriques19, DeVisser21, Bargerbos23}, substrate etching to block phonon propagation~\cite{Chu16, Rostem18, Karatsu19}, backside metallization to enhance phonon downconversion~\cite{Iaia22, Yelton24, Larson25}, QP traps using normal and low-gap superconductors~\cite{Riwar16, Riwar19, Martinis21, Pan22, Thorbeck23}, and band-gap engineering of JJ electrodes~\cite{Diamond22, kamenov23, Connolly24, McEwen24, kurilovich25, Nho26}. These innovations are often assessed not by their impact on the rate or recovery dynamics of qubit error bursts but on the average QP density in qubit JJ electrodes, extracted by measuring QP tunneling rates~\cite{Riste13}. Some experiments observe exponential or power-law time dependence of QP tunneling rates after cooldown~\cite{Mannila22, Anthony24, Yelton25, kerschbaum26}, attributed to gradual thermalization of substrates, metal films, device packaging, and shielding.

In this Letter, we investigate qubit error bursts in two transmon-based processors of Quantum Inspire~\cite{QI_website}, Starmon-5 (henceforth $\Sfive$) and Starmon-7 ($\Sseven$), with similar design, fabrication and packaging but different Al/AlOx/Al JJ types (Dolan~\cite{Dolan77} versus Manhattan~\cite{Potts01}, respectively). We extract the rate (with and without lead shielding) and duration of bursts in each device in two dilution refrigerators (A and B) to discern device-specific from refrigerator-dependent characteristics. The rate and duration of error bursts are device specific and within the range reported by prior experiments. However, we observe two unforeseen signatures only in $\Sfive$. First, the duration of bursts is shortened by increasing the rate of qubit $\pi$ pulsing in two variants of the detection scheme. We explain this effect as a pulse-induced QP pumping mechanism possible only in the Dolan-bridge JJs, whose fabrication produces a built-in QP trap. The second signature is an anomalous time dependence in the error burst rate: after holding steady for days or weeks since cooldown, the rate  surges by one to two orders of magnitude for several hours and then gradually decreases over days to a value two orders of magnitude lower than immediately post-cooldown, persisting until a thermal cycle. The physical mechanism for this anomalous time dependence remains unclear, inviting suggestions. The vanishing of the burst rate post-surge could be exploited to mitigate radiation-induced correlated errors.

\begin{figure}[ht]
\centering
\includegraphics[width=\columnwidth]{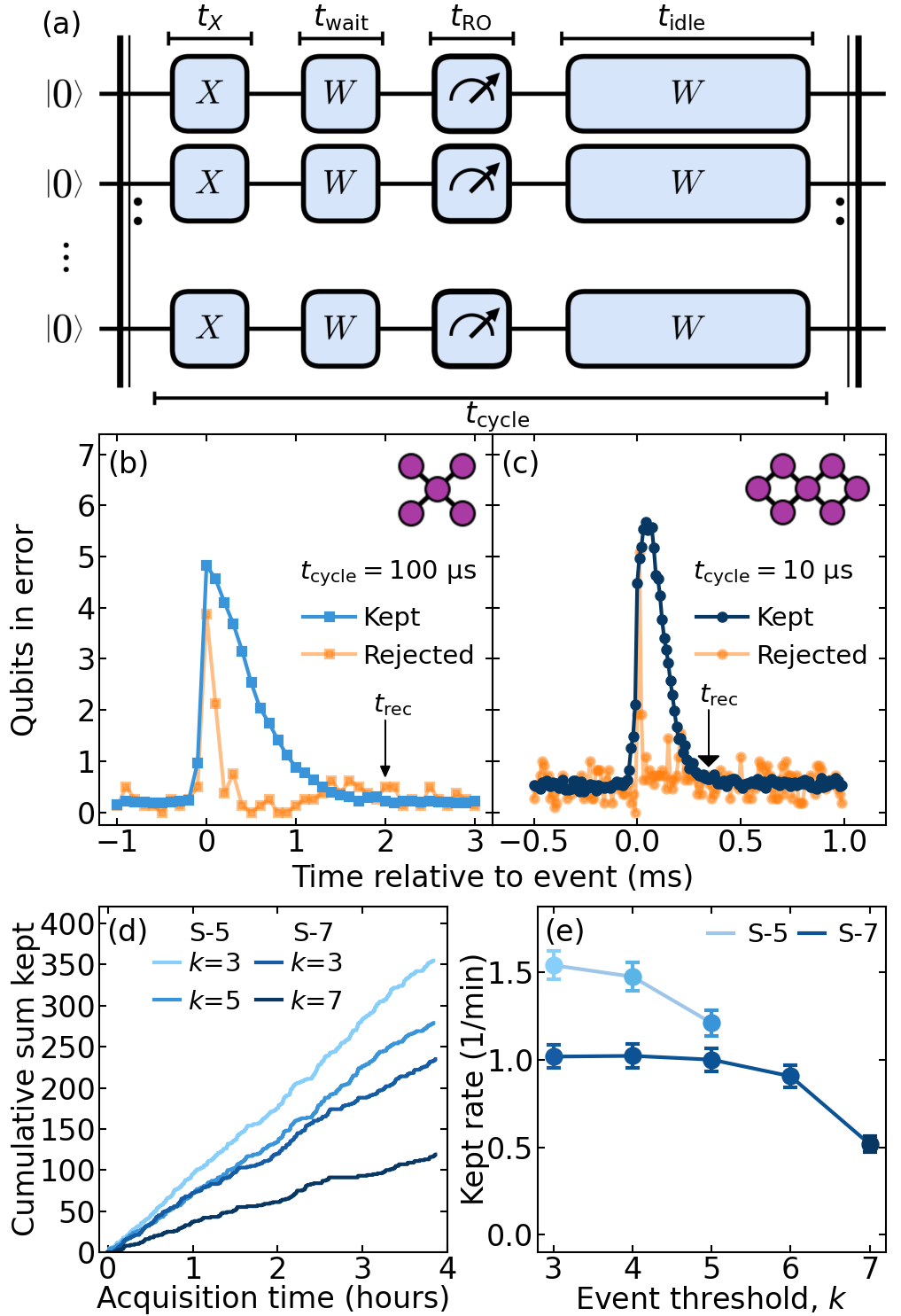}
\caption{
Detection and characterization of simultaneous error events.
(a) Repeated detection subcircuit described in the main text. Blocks labelled $W$ correspond to idling (waiting) periods.
(b,c) Average of kept and rejected events for (b) $\Sfive$ ($\tcycle=100~\us$, $\nth=5$) and (c) $\Sseven$
($\tcycle=10~\us$, $\nth=7$). Note the different timescales on horizontal axes. For kept events in $\Sfive$ ($\Sseven$), the baseline is fully recovered in $\trec \sim 2~\ms$ $(\sim250~\us)$.
(d) Cumulative sum of kept events for $\Sfive$ ($\Sseven$) for $\nth=3$ and maximized $\nth=5$ $(7)$.
(e) Extracted rate of kept events $\gammakept$ as a function of $\nth$ for both processors.
}
\label{fig:Figure1}
\end{figure}

The burst detection scheme [Fig.~\ref{fig:Figure1}(a)] consists of back-to-back repetitions of the following subcircuit: all qubits are $\pi$ pulsed in $\tx=20~\ns$, idled for $\twait=2~\us$, readout in $\tro$, and
idled again for $\tidle$, for a total subcircuit duration
$\tcycle = \tx + \twait + \tro + \tidle$ [$\tro=2~(1)~\us$ in $\Sfive~(\Sseven)$]. Later, we vary $\tcycle$ through $\tidle$.
The string of measurement outcomes for each qubit is ideally a series of alternating 0s and 1s.
We define a qubit error as two consecutive 0 outcomes.
We do not consider two consecutive 1 outcomes as an error because leakage to the second-excited state, which can be residually induced by measurement, produces this signature as well. We define a simultaneous error event whenever the number of qubits in error $n$ peaks at $n\geq \nth$, with $\nth$ a set event threshold.

To distinguish bursts originating from ionizing radiation from error events caused by the baseline qubit relaxation time $\Tone$ and readout ($\RO$) errors~\cite{SOM_CR}, we post-process the time series of $n$ using two passes.
The first marks all simultaneous events using the $n\geq \nth$ condition. The second pass separates events based on their duration.
For bursts induced by ionizing radiation, the characteristic time $\trec$ for $n$ to fully return to the baseline varies widely across reported experiments,
from $\sim 150~\us$~\cite{Li24} to $\sim 100~\ms$~\cite{McEwen22}. For error events triggered by baseline $\Tone$ and $\RO$ errors, the baseline is recovered in just $\sim 2 \tcycle$. Therefore, a large separation of scales between $\tcycle$ and $\trec$ is best for discrimination.
The second pass uses template matching following Refs.~\cite{McEwen22, Wu25}, with a stepped exponential decay template function (see~\cite{SOM_CR}).
We use the peak value of the matched-filter output to set the start ($t=0$) of each event, which is used when averaging over events.
We manually optimize the template function decay constant to best separate the two apparent distributions in the histogram of peak output values.
As $\nth$ increases, the edge of the distribution associated with events caused by baseline $\Tone$ and RO errors (which we aim to reject) shifts to lower values [see Fig.~\ref{fig:cr_filtering}(d) in~\cite{SOM_CR}]. We therefore set the threshold for keeping and rejecting events at $\nth=3$ to minimize false positives, and apply this same threshold for all cases with $\nth>3$.

We first estimate $\trec$ in the two processors by averaging the time series of $n$ for kept events.  For $\Sfive$~$(\Sseven$), we use $\tcycle=100~(10)~\us$ and set $\nth=5~(7)$ to maximize signal range. We find $\trec\sim2~\ms$ for $\Sfive$ and $\sim250~\us$ for $\Sseven$. This $\times8$ difference is device specific, as confirmed by repeating measurements for each device in refrigerator B (see Fig.~\ref{fig:FigKeptFridges} in~\cite{SOM_CR}). Furthermore, we have also found that $\trec$ is robust to variations in packaging. For example, doubling the density of ground-plane wirebonds in $\Sseven$ to roughly match that in $\Sfive$ did not further shorten its $\trec$. In $\Sfive$, adding small amounts of GE varnish to the backside corners of the substrate to improve thermal contact and clamping (which $\Sseven$ has) also did not shorten its $\trec$. We speculate that the difference in $\trec$ between the processors may instead be rooted in geometrical differences between their JJ electrodes (see~\cite{SOM_CR} for images). Effective QP trapping and recombination rates vary by more than one order of magnitude in the literature (see~\cite{Wang14, Derooij21} and references therein) and are known to depend on electrode geometry. Detailed modeling beyond our present capabilities is required to substantiate this.

\begin{figure}[ht]
\centering
\includegraphics[width=\columnwidth]{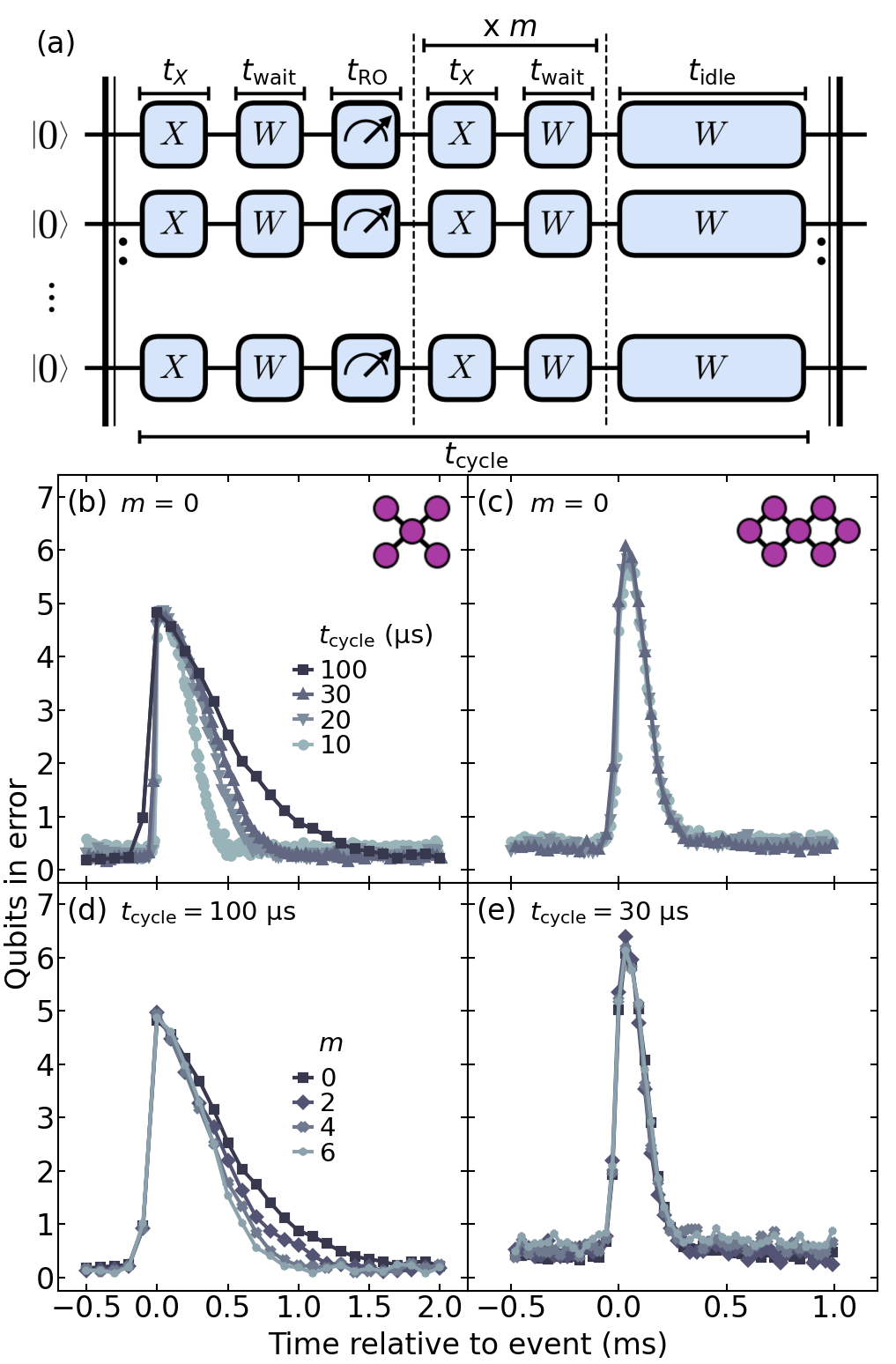}
\caption{
Signature of QP pumping in $\Sfive$ and absence thereof in $\Sseven$.
(a) Variant of the repeated detection subcircuit: an even number $m$ of $\pi$ pulses is inserted after readout.
(b,c) Average of kept events for (b) $\Sfive$ and (c) $\Sseven$ with fixed $m=0$ and varying $\tcycle$. $\Sfive$ shows an increase in $\trec$ as $\tcycle$
increases from $10$ to $100~\us$. $\Sseven$ does not show any such dependence.
(d,e) Average of kept events for (c) $\Sfive$ and (d) $\Sseven$ with fixed $\tcycle=100$ and $30~\us$, respectively, but increasing $m$.
In $\Sfive$, but not in $\Sseven$, $\trec$ decreases as $m$ increases from 0 to 6. In combination, these results demonstrate that increasing the net rate of $\pi$ pulses
in the detection scheme shortens $\trec$ in $\Sfive$.
}
\label{fig:Figure2}
\end{figure}

\begin{figure}[ht]
\centering
\includegraphics[width=\columnwidth]{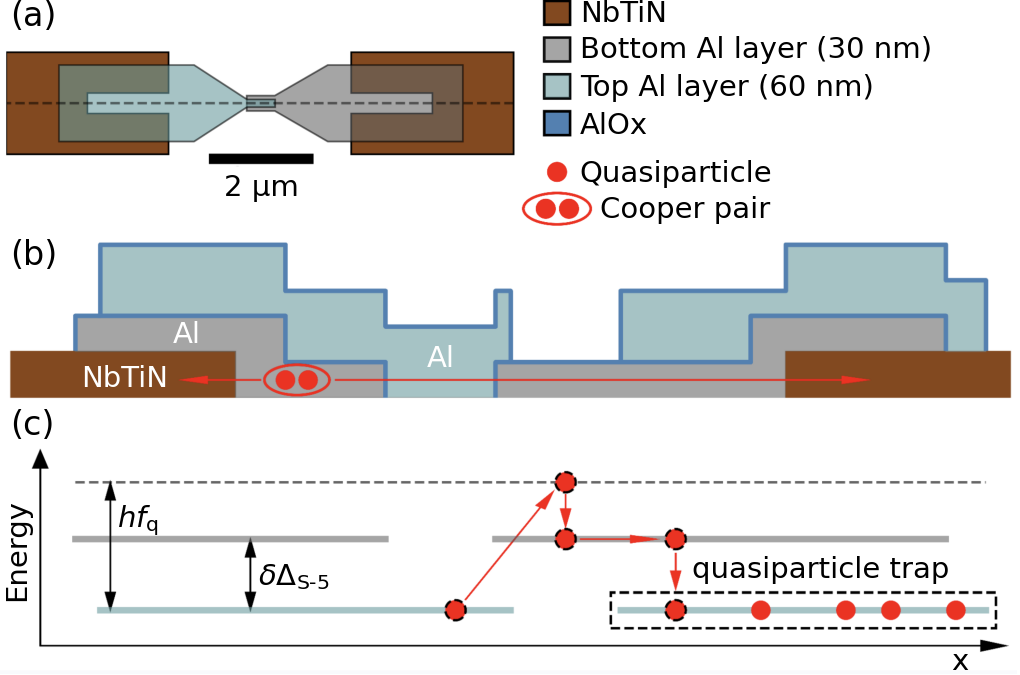}
\caption{
Schematics of the transmon Josephson junctions of $\Sfive$ fabricated using the Dolan-bridge technique.
(a) Top view. (b) Cross section along the long axis. (c) Energy diagram illustrating the superconducting energy gaps $\DeltaAl$ in the thin (bottom) and
thick (top) Al layers deposited by double-angle evaporation. The difference between these gaps $\delta \DeltaSfive$ is smaller than the qubit transition energy $h \fq$ in each transmon.
[The much larger gap of the NbTiN transmon capacitor pads $(\Delta_{\mathrm{NbTiN}}\sim2.3~\mathrm{meV})$ is not shown.]  With the qubit in $\ket{1}$, a QP on the thick left layer can tunnel across the dominant JJ into the thin right layer by relaxing the qubit to $\ket{0}$, then diffuse and tunnel onto the thick right layer, where it becomes trapped. A subsequent $\pi$ pulse re-excites the qubit, commencing another cycle of QP pumping to the trap.
}\label{fig:Figure3}
\end{figure}

Having characterized their recovery time, we next turn to the frequency of bursts. The rate of kept events $\gammakept$ is extracted from $\sim 4~\hours$ of acquisition time (not wall-clock time) for each processor.
In $\Sfive$, we observe $\gammakept = 1/(38.9\pm2.1~\s)$ at $\nth=3$ [Fig.~\ref{fig:Figure1}(d)]. We can compare this rate to the known muon flux at sea level~\cite{Cecchini12}, approximately $1~\cm^{-2}\minutes^{-1}$, and thus $1/(90~\s)$ for the $8~\mm \times 8~\mm$ area common to both processors.
In $\Sseven$, we observe $\gammakept = 1/(58.9\pm3.8~\s)$ at $k=3$ [Fig.~\ref{fig:Figure1}(e)]. It is possible that the lower $\gammakept$ may be due to a higher rate of false rejects in $\Sseven$ resulting from the short $\trec$. We can compare these rates to reported values in experiments~\cite{McEwen22, Harrington24, McEwen24} that use similar detection schemes. Accounting for device dimensions (see~\cref{tab:TableRates}) shows  that $\gammakept$ in $\Sfive$ is comparable to that in Ref.~\cite{Harrington24},
while $\gammakept$ in $\Sseven$ is comparable to those in Refs.~\cite{McEwen22, McEwen24}. Figure~1(e) shows the dependence of $\gammakept$ on $\nth$, revealing that the majority of bursts affect the majority of qubits in each device.

Varying $\tcycle$ reveals a more striking difference between $\Sfive$ and $\Sseven$ that connects to the different types of JJ used for their transmons (Fig.~\ref{fig:Figure2}). In $\Sfive$, varying $\tcycle$ from $100$ to $10~\us$ does not affect $\gammakept~(\nth=3)$~\cite{SOM_CR}, but decreases $\trec$ from $\sim2~\ms$ to $\sim500~\us$. In sharp contrast, no change in $\trec$ is
observed in $\Sseven$ over the same $\tcycle$ range (no data shown for $\tcycle=100~\us$).
A similar effect is observed in a variant of the detection subcircuit where, keeping $\tcycle=100~(30)~\us$ for $\Sfive$ $(\Sseven)$, we insert an even number $m$ of $\pi$ pulses during $\tidle$. Again, $\trec$ decreases monotonically with increasing $m$ for $\Sfive$, but not $\Sseven$.

\begin{figure*}[ht]
\centering
\includegraphics[width=\textwidth]{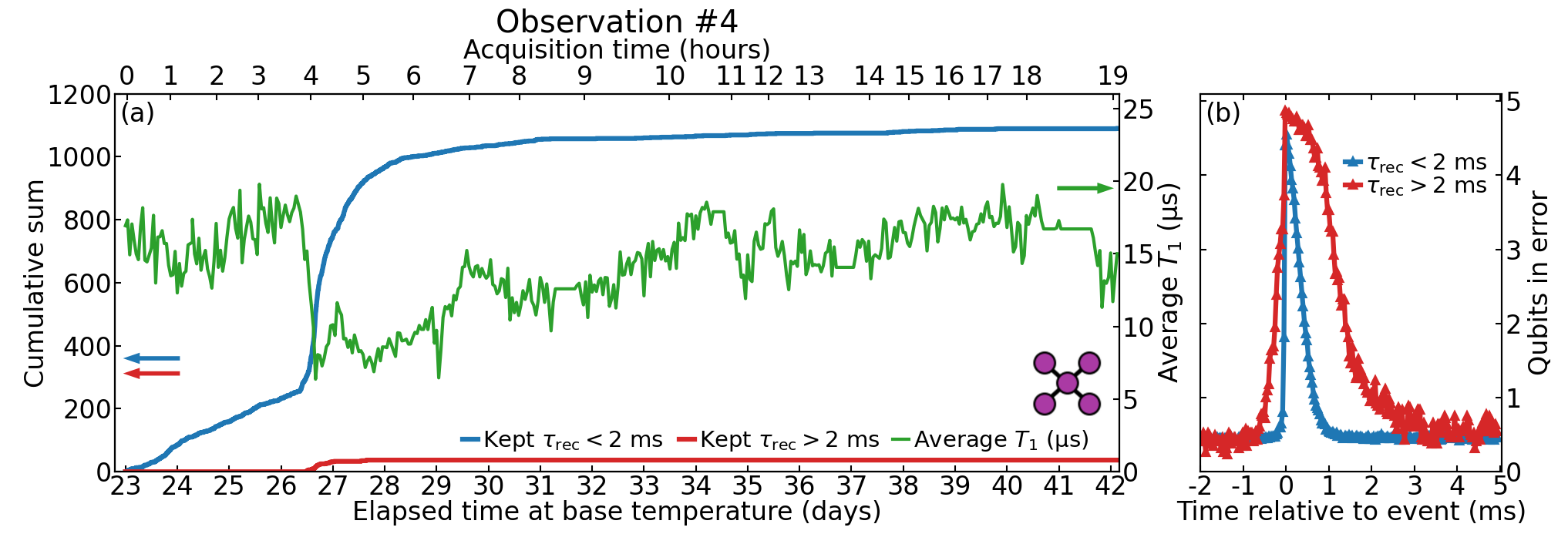}
\caption{
Example observation of the surge in $\Sfive$ while in refrigerator A.
(a) Cumulative sum of kept events (left axis) and average $\Tone$ (right) as a function of elapsed time at base temperature (bottom) and elapsed time running the detection circuit of Fig.~\ref{fig:Figure1}(a) ($\tcycle=30~\us$, $\nth=3$).
After 26 days at base temperature with similar characteristics as Fig.~\ref{fig:Figure1}, $\gammakept$ suddenly surges by a factor $\sim\times10$ and average $\Tone$ drops by $\sim\times3$. Over the next several days, $\gammakept$ monotonically decreases, ultimately settling to a value $\sim \times 100$ lower than at the start of the experiment, and average $\Tone$ gradually recovers.
(b) Average of kept events ($\tcycle=30~\us$, $\nth=5$) over the full observation period.
We use a second pass of template matching (see~\cite{SOM_CR}) to discern $37$ bursts with atypically long recovery times (red), all of which are detected at the onset of the surge. See~\cite{SOM_CR} for other observations of the surge showing some refrigerator-specific characteristics.
See~\cite{SOM_CR} for other observations of the surge in $\Sfive$.
}
\label{fig:Figure4}
\end{figure*}

To understand why the net rate of $\pi$ pulses in the detection scheme affects $\trec$ in $\Sfive$, recall that a prior experiment~\cite{Gustavsson16} has shown active reduction of QP densities near active qubit JJs through repeated $\pi$ pulsing. QP pumping, as the effect is known, was first evidenced as pulse-controlled enhancement of $\Tone$ in a flux qubit.
Here, the effect is evidenced as a pulse-controlled reduction of $\trec$ in $\Sfive$. The reason becomes clear upon noting that Dolan-bridge JJs like those of $\Sfive$ have a built-in QP trap (Fig.~\ref{fig:Figure3}), while
Manhattan-style JJs like those of $\Sseven$ do not (see Fig.~\ref{fig:FigDevices}).

Our Dolan JJs are defined by double-angle evaporation of Al layers with intermediate oxidation~\cite{Muthusubramanian24}, with bottom (top)
layer thickness $d=30~(60)~\nm$. The cross-section schematic in Fig.~\ref{fig:Figure3}(b) shows the dominant active JJ near the center, which has the smallest overlap area
and is largely responsible for setting the qubit transition frequency $\fq$. There are two more JJs  created by the shadowing mechanism: a large active one to the left and another large inactive one to the right.

Owing to their different $d$, the bottom and top layers are expected to have different superconducting gap $\DeltaAl$.
For thin Al films, $\DeltaAl$ increases with decreasing $d$. Using the empirical model~\cite{Marchegiani22}
\[
\DeltaAl(d) = \DeltaBulk + \alpha/d
\]
with $\DeltaBulk = 180~\ueV$ and $\alpha = 600~\ueV \cdot \nm$, we estimate a gap difference between the layers
$\delta \DeltaSfive \sim h \times 2.4~\GHz$. As $h \fq > \delta \DeltaSfive$ for all its  transmons, we do not expect band-gap engineering effects~\cite{Diamond22, kamenov23, Connolly24, McEwen24, kurilovich25, Nho26} to play a role in $\Sfive$.

During a burst, if the qubit is in the first-excited state $(\ket{1})$, a QP at the band edge of the thick layer on the left can tunnel across the dominant JJ
onto the thin layer on the right by absorbing the qubit energy (i.e., relaxing the qubit to the ground state $\ket{0}$). The QP can then diffuse rightward, tunnel across to the thick layer on the right, and drop to the band edge, becoming trapped [Fig.~\ref{fig:Figure3}(c)].  In this way, repeated qubit re-excitation by $\pi$ pulses creates a net pumping of QPs away from the thick layer to the left of the dominant JJ to the trap on the right. This pumping process shortens $\trec$.

Finally, we report another phenomenon observed only in $\Sfive$ but for which we currently lack explanation, inviting suggestions from the community.
The phenomenon is an anomalous time dependence in $\gammakept$ that is best described as a surge. We do not find any link between this surge and any sudden change in any pressure or temperature within the refrigerator or any applied microwave-frequency or baseband pulsing. Figure~\ref{fig:Figure4} presents a typical observation in refrigerator A
for $\tcycle=30~\us$.
After 3-4 weeks at base temperature showing a steady value, $\gammakept$ surges by an order of magnitude, accompanied by
a sharp decrease in average $\Tone$. During this surge, some kept events show uncharacteristically long $\trec~\sim3~\ms$. Over the next few days, $\gammakept$ decreases
monotonically and ultimately settles to a value $\sim\times100$ lower than at the start of the experiment. In parallel, the average $\Tone$ gradually recovers to pre-surge values.
The almost vanishing $\gammakept$ persists until thermal cycling. We have performed cycles to $4~\K$ as well as to room temperature, finding that $\gammakept$ fully resets
in both cases.
The surge observed in $\Sfive$ has some refrigerator-specific characteristics. In refrigerator A, the surge has occured in all 4 cooldowns since it was first identified,
with the characteristics described above (see~\cite{SOM_CR} for more observations). In refrigerator B, the surge has been observed in $12$ out of $20$ cooldowns. In 11 of
them, the surge began just $\sim1$ day after reaching base temperature, and after $\sim1$ week in the other. The onset is marked by an increase in $\gammakept$ by
$\sim\times100$, i.e., even stronger than in refrigerator A. Cooldowns during which the surge did not occur have maintained base temperature for periods ranging between
$4$ and $42$ days. Curiously, the surge has not been observed in any of 4 cooldowns since adding GE varnish to the backside of $\Sfive$. This is noteworthy as $\Sseven$,
which has had such GE varnish from the beginning, has not shown any surge in either refrigerator to date. However, more cooldowns are required to establish
causality.

Equally intriguing as the trigger source for the surge is the physical mechanism causing $\gammakept$ to practically vanish in the days following. Evidently, no significant time dependence is known to occur in the incidence rate of ionizing radiation of cosmogenic origin (or local radioactivity in our lab). Rather, some on-chip mechanism must be at play.


In summary, we have investigated qubit error bursts in the 5- and 7-transmon processors made available (at different times) via the Quantum Inspire platform.
We use a detection scheme exploiting mid-circuit measurement and signal-processing passes to discern real bursts from false positives due to baseline $\Tone$ and readout errors, despite the low qubit count.
$\Sfive$ and $\Sseven$ exhibit refrigerator-independent burst rates $\gammakept \sim 1.5$ and $1~\minutes^{-1}$, respectively, comparable to those reported in Refs.~\cite{McEwen22, Harrington24, McEwen24} after accounting for device dimensions.
The processors reveal more differences that are interesting in view of the similarity in processor design, fabrication and packaging.
First, the recovery time of bursts is also refrigerator independent and differs by a factor $\times 8$, possibly due to difference in JJ electrode geometries. Second,  $\trec$ in $\Sfive$ can be shortened by increasing the rate of applied $\pi$ pulses in the detection scheme. This is a manifestation of QP pumping away from the transmon JJ into a nearby QP trap created by the Dolan-bridge JJ fabrication process. Such pumping is not possible in $\Sseven$ as its Manhattan-style JJs lack such built-in trap. Third, $\Sfive$ shows an anomalous surge and subsequent reduction of $\gammakept$, with some refrigerator-dependent characteristics. Given the obvious benefit to QEC of limiting the rate, duration and globality of qubit error bursts, understanding the physics of the surge and its aftermath motivates further research.

\section{Acknowledgements}
We thank J.~Baselmans, K.~Karatsu, C.~Andersen, B.~Plourde, L.~Kouwenhoven, G.~Frossati, and their teams for helpful discussions; S.~van der Meer and M.~Sarsby for assistance with dilution
refrigerators; G.~Calusine and W.~Oliver for providing traveling-wave parametric amplifiers; the TU Delft Reactor Institute for the lead shield;
the Kavli Nanolab Delft staff for cleanroom support; and V.~Sinha for project management assistance.
This research is funded by the Dutch National Growth Funds and Quantum Delta NL (KAT-1 StartImpulse and Phase 2), the European Union Flagship on Quantum Technology (OpenSuperQplus100, No. 101113946), Intel Corporation,  the Dutch Ministry of Economic Affairs [Allowance for Top Consortia for Knowledge and Innovation (TKI)], and the Netherlands Organization for Scientific Research (NWA.1292.19.194). M.S. acknowledges funding from the Google PhD fellowship program.

\section{Authors Contribution}
G.R.D.C., M.S., J.F.M., and L.D.C conceptualized the experiment.
G.R.D.C. and M.S. performed the experiment and data analysis.
J.F.M. performed initial experiments.
N.M. fabricated $\Sfive$, A.K. and M.F. fabricated $\Sseven$.
M.B., N.H. and L.D.C. designed the two processors.
M.S.M. implemented mid-circuit measurement and B.S. developed monitoring software in Quantum Inspire.
S.V.S. performed post-fabrication trimming of $\Sseven$ resonators.
G.R.D.C., M.S. and L.D.C. wrote the manuscript with feedback from coauthors.
L.D.C. supervised the project.
G.R.D.C. and M.S. contributed equally to this work.

\section{Competing Interests}
The authors declare no competing interests.

\section{Data Availability}
All data shown are available at \url{https://github.com/DiCarloLab-Delft/QubitErrorBursts}.
Further materials can be provided upon request. 

%

\end{bibunit}

\onecolumngrid
\clearpage

\renewcommand{\theHfigure}{S\arabic{figure}}
\renewcommand{\theHequation}{S\arabic{equation}}
\renewcommand{\theHtable}{S\arabic{table}}

\renewcommand{\theequation}{S\arabic{equation}}
\renewcommand{\thefigure}{S\arabic{figure}}
\renewcommand{\thetable}{S\arabic{table}}
\renewcommand{\bibnumfmt}[1]{[S#1]}
\renewcommand{\citenumfont}[1]{S#1}
\setcounter{figure}{0}
\setcounter{equation}{0}
\setcounter{table}{0}

\newcommand{\mK}{\mathrm{mK}}
\newcommand{\mtr}{\mathrm{m}}
\newcommand{\myday}{\mathrm{d}}

\begin{bibunit}[apsrev4-2]
\section*{Supplementary Material for 'Qubit error bursts in superconducting quantum processors of Quantum Inspire: quasiparticle
pumping and anomalous time dependence'}

\twocolumngrid

This supplement provides additional information in support of statements and claims made in the main text.

\section{Characteristics of the two quantum processors}
$\Sfive$ and $\Sseven$ have many features in common but also noteworthy differences.
Both have 7 transmons with nearest-neighbor connectivity (using fixed-frequency bus resonators) originally intended for the distance-2 surface code~\cite{Marques22}. All transmons have dedicated microwave drive lines for single-qubit gates, flux-control lines for DC flux biasing at their sweetspot and two-qubit gates  (not used here), and dedicated readout and Purcell-filter resonator pairs enabling frequency-multiplexed readout using two feedlines.
While the topology is the same for both processors, there is an evolution in design and layout from $\Sfive$  (fabricated in 2019) to $\Sseven$ (2024).
For example, $\Sseven$ has 'shoelacing' airbridges allowing  post-fabrication frequency trimming of readout and Purcell resonators~\cite{Valles23}.
Improved frequency matching of readout and Purcell resonator pairs enables readout in $\Sseven$ to be faster.
For simplicity, $\Sseven$ does not have in-line capacitors at the feedline inputs.

$\Sfive$ is so named because in actuality two of its transmons are not operable.
One has a defect in the base layer that shorts one of its capacitor pads to the ground plane.
The other cannot be measured as its readout resonator cannot even be identified in spectroscopy (cause unknown, but likely to be large detuning from the Purcell resonator).

Both processors are $8~\mm \times 8~\mm$ dies, with NbTiN base layers (thickness $200~\nm$) sputtered on high-resistivity Si substrates (100 orientation, thickness $525~\um$).
However, the base layers were sputtered using different systems (SuperAJA for $\Sfive$ and Nordiko for $\Sseven$).
Also, the substrate of $\Sfive$ is single-side polished while that of $\Sseven$ is double-side polished.
All circuit elements on the base layer, including the transmon capacitor pads, are defined by electron-beam lithography and reactive-ion etching, and wet echting in an ammonia-peroxide mixture.

\begin{figure*}[ht]
\includegraphics[width=0.8\textwidth]{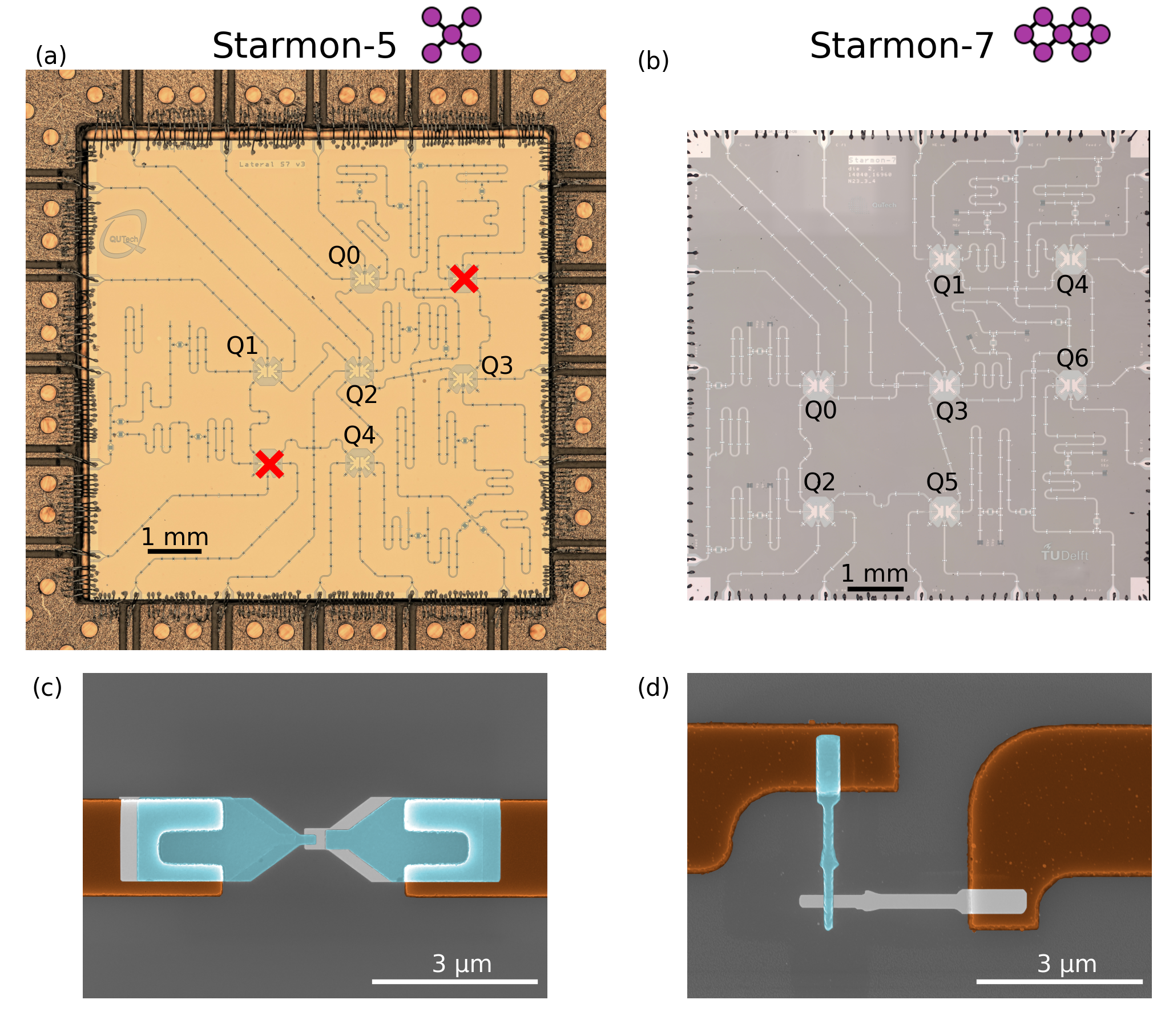}
\caption{
(a,b) Optical images of the (a) $\Sfive$ and (b) $\Sseven$ quantum processors. Both have 7 transmons and the same nearest-neighbor connectivity, with only evolutionary differences in design and layout (e.g., signal routing). Two transmons of $\Sfive$ are inoperable (indicated by red crosses). A key addition in $\Sseven$ is the addition of 'shoelacing' airbridges allowing post-fabrication frequency trimming of readout and Purcell resonators~\cite{Valles23}. Both processors are laterally connected to a Cu PCB using Al wirebonds. These images were taken before the start of the experiment, when the initial density of wirebonds in $\Sfive$ was roughly double that of $\Sseven$.
(c,d) Scanning electron microscope images of (c) Dolan and (d) Manhattan JJs in sister devices of $\Sfive$ and $\Sseven$, respectively, with added falsecolor: gray for the bottom (thin) electrode, light blue for the top (thick) electrode, and brown for the NbTiN arms of the superconducting quantum interference device (SQUID) loops.  Note that we do not use any bandaging layers between the Al electrodes and the NbTiN arms.}
\label{fig:FigDevices}
\end{figure*}

An important fabrication difference between the processors lies in their Al/AlOx/Al JJs.
$\Sfive$ has Dolan-bridge JJs, while $\Sseven$ has Manhattan-style JJs.
Scanning electron microscope images of the JJs of devices fabricated in the same batch as $\Sfive$ and $\Sseven$ are shown in Figs.~\ref{fig:FigDevices}(c) and~\ref{fig:FigDevices}(d), respectively. Importantly, note  the absence of extra inactive JJs in the electrodes of $\Sseven$. With bottom (top) layer thickness $d=30~(140)~\nm$, we estimate $\delta \DeltaSseven \sim h \times 3.8~\GHz < h \fq $ for all transmons. Therefore, we also do not expect band-gap engineering effects to play a role in $\Sseven$.

Both processors have Al airbridges and crossovers to achieve the required connectivity and signal routing.
Both are connected to identical 20-port Cu printed circuit boards (PCBs) using  AlSi wirebonds.
Each PCB is encapsulated by an Au-plated Cu lid and base, the latter with a vacuum pocket to raise the frequency of package resonance modes. Together, the lid and base form the packaging, which always accompanies the device when changing dilution refrigerator.

There were two more noteworthy differences at the start of the experiment.
First, the density of wirebonds in $\Sfive$ was roughly double that in $\Sseven$ (evident in Figs.~\ref{fig:FigDevices}(a) and \ref{fig:FigDevices}(b), respectively).
Second, $\Sseven$ had small droplets of GE varnish on its backside corners, used to fix the chip to the package base and thereby avoid displacement during wirebonding.
$\Sfive$ was packaged and wirebonded without any varnish.

Later on, we made several modifications to test if any of them might affect $\trec$ and $\gammakept$.
In $\Sfive$, we removed all of the original wirebonds as we found some to be loosely connected or even fully detached. Such imperfections were avoided when re-wirebonding at similar density.
Later on, we also added small droplets of GE varnish to its backside corners.
For $\Sseven$, we also added wirebonds to make the density of ground-plane wirebonds similar to that of $\Sfive$.
As noted in the main text (data not shown), none of these variations affected $\trec$ or $\gammakept$ in either processor.
However, the surge has not been observed in $\Sfive$ in the four cooldowns following the addition of GE varnish.

A summary of typical operational parameters for both processors is provided in~\cref{tab:TableS5,tab:TableS7}.

\section{Characteristics of the two dilution refrigerators}
Refrigerators A (Leiden Cryogenics MNK 650 CF) and B (Leiden Cryogenics CF-1000) are $~2.7~\mtr$ apart without any intervening wall.
The refrigerators have many common features but also noteworty differences.
They cool from room temperature (RT) to a mixing-chamber (MC) base temperature $\sim 20~\mK$ in $\sim 1$~day using liquid-nitrogen precooling lines.
They have Al shields at the RT, $50~\K$ and $4~\K$ plates, without any internal coating for infrared (IR) absorption.
Furthermore, both refrigerators have Au-plated Cu shields at the Still, cold plate (CP), and MC. In refrigerator A, only the MC shield is internally coated with Aeroglaze for IR absorption. In refrigerator B, the three shields have such coating.

In refrigerator A, there is one outer can of Cryophy magnetic shielding, and a double-walled inner can with Al outer and Cu inner wall.
The inner wall is coated with a layer of SiC powder mixed into Stycast~2850 for IR absorption.
In refrigerator B, there are two outer cans of Cryophy magnetic shielding, and a double-walled inner can with Al outer and Cu inner walls.
The inner wall is Au plated and coated with a layer of SiC powder mixed into Stycast~2850 for IR absorption.

\begin{figure*}[ht]
\includegraphics[width = 0.9\textwidth]{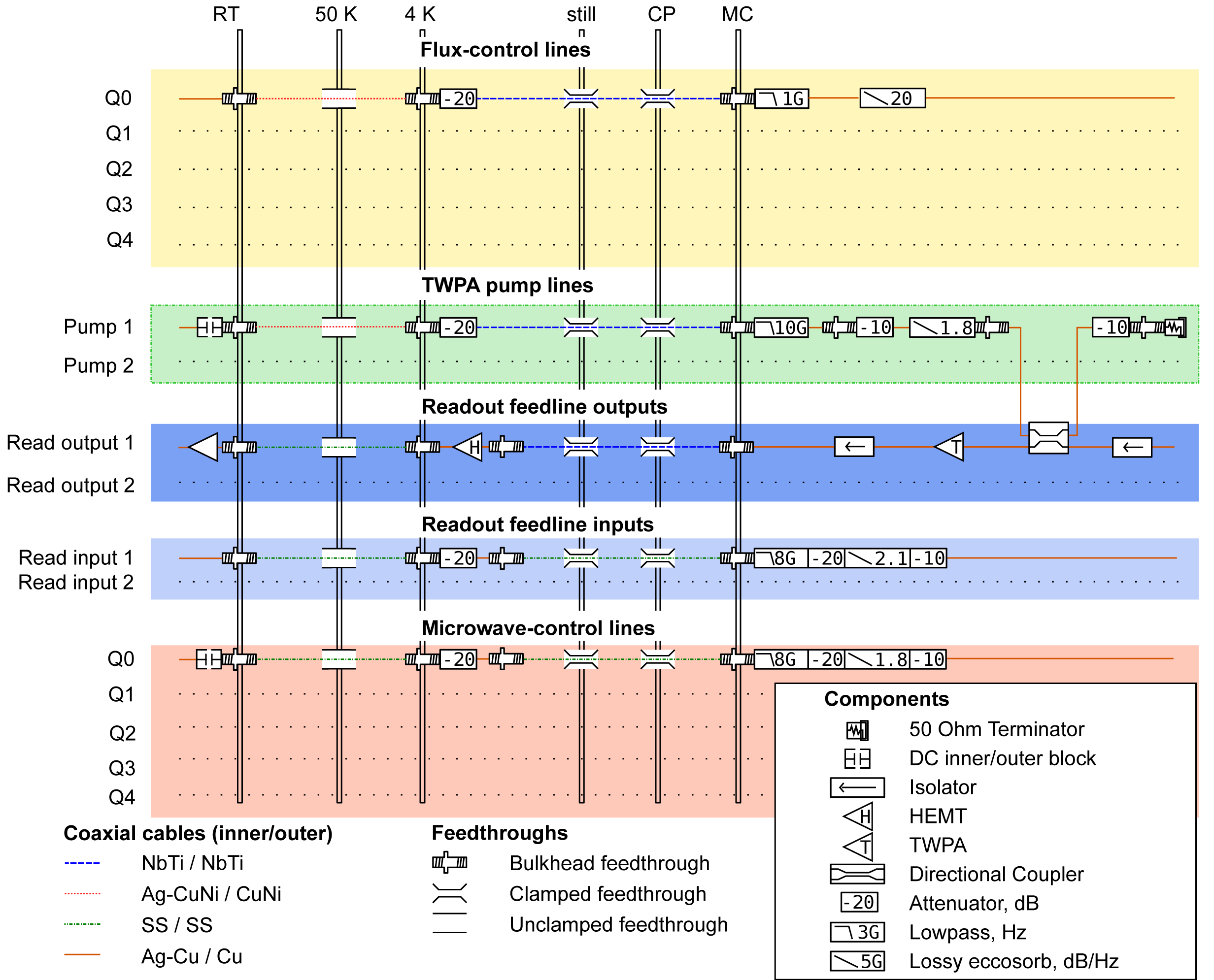}
\caption{
Wiring diagram of refrigerator A for $\Sfive$.
}
\label{fig:FigWiring}
\end{figure*}

The refrigerators are similarly wired using UT-85 semirigid coaxial cables between RT and MC, hand-formable (also UT-85-size) coaxial cables at the MC, and standard attenuation and filtering profiles.
In both refrigerators, the readout amplification chain for each feedline (two in each processor) consists of a travelling-wave parametric amplifier (TWPA) located at the MC,
a high-electron-mobility-transistor (HEMT) amplifier at $4~\K$, and a MITEQ AFS-3 amplifier at RT.
For reference, a detailed schematic of the wiring of refrigerator A is shown in Fig.~\ref{fig:FigWiring}.

\begin{table*}[!ht]
\centering
\caption{
Typical operational characteristics of $\Sfive$ in refrigerator A (comparable to those in refrigerator B).
Coherence times are extracted from standard time-domain experiments~\cite{Krantz19} performed prior to any surge.
}
\label{tab:TableS5}
\begin{tabularx}{\textwidth}{p{10cm}  p{1.0cm} p{1.0cm} p{1.0cm} p{1.0cm} p{1.0cm} p{1.0cm}}
\hline
$\Sfive$ transmon & $\text{Q0}$ & $\text{Q1}$ & $\text{Q2}$ & $\text{Q3}$ & $\text{Q4}$ & Average\\
\hline
Qubit frequency at bias point (sweetspot), $\fq~(\GHz)$   & 6.425     & 6.517     & 5.695     & 5.047     & 4.876    &\\
Transmon anharmonicity, $\alpha/2\pi~(\MHz)$                                & -265      & -260      & -285      & -300      & -295      &\\
Readout frequency, $\fRO~(\GHz)$                        & 7.905     & 7.838     & 7.439     & 7.298     & 7.126     &\\
Relaxation time, $\Tone~(\us)$                                              & 15        & 21        & 19        & 23        & 15        & 18.6\\
Ramsey dephasing time, $\TtwoRamsey~(\us)$                                  & 9         & 12        & 15        & 14        & 8         & 11.6\\
Echo dephasing time, $\Ttwoecho~(\us)$                                      & 27        & 37        & 33        & 43        & 16        & 31.2\\
Multiplexed average assignment fidelity, $F_{\text{RO}}~(\%)$               & 97.0      & 98.4      & 98.3      & 98.7      & 97.9      & 98.1\\
\hline
\end{tabularx}
\end{table*}

\begin{table*}[!ht]
    \centering
\caption{
Typical operational characteristics of $\Sseven$ in refrigerator A. For the device parameters listed here we have observed
a three-transmon interaction between qubits Q1, Q3 and Q4 occuring only when preparing all three in $\ket{1}$ and measuring them simultaneously,
resulting in some leakage in Q1 and relaxation in Q3 and Q4. The effect diminishes when biasing Q3 and Q4
$\sim 200~\MHz$ away from their sweetspot; repeating the circuit variant shown in Fig~\ref{fig:Figure1}(a) in this configuration showed no quantitative
difference in both the averaged response to kept simultaneous error events and $\gammakept$ (data not shown).
}
\label{tab:TableS7}
\begin{tabularx}{\textwidth}{p{8.5cm}  p{0.9cm} p{0.9cm} p{0.9cm} p{0.9cm} p{0.9cm} p{0.9cm} p{0.9cm} p{0.9cm}}
\hline
$\Sseven$ transmon & $\text{Q0}$ & $\text{Q1}$ & $\text{Q2}$ & $\text{Q3}$ & $\text{Q4}$ & $\text{Q5}$ & $\text{Q6}$ & Average\\
\hline
Qubit frequency at bias point (sweetspot), $\fq~(\GHz)$      & 6.251     & 6.227     & 5.590     & 5.757 & 5.753 & 4.484     & 4.659     &\\
Transmon anharmonicity, $\alpha/2\pi~(\MHz)$                        & -310      & -315      & -310      & -315      & -310  & -330      & -315              &\\
Readout frequency, $\fRO~(\GHz)$               & 7.571     & 7.481     & 7.145     & 7.116     & 7.265 & 6.958     & 6.616             &\\
Relaxation time, $\Tone~(\us)$                                      & 31        & 15        & 23        & 18        & 22    & 36        & 49                & 27.7\\
Ramsey dephasing time, $\TtwoRamsey~(\us)$                          & 11         & 7         &  5        & 11         & 13    & 2         & 8               & 8.1\\
Echo dephasing time, $\Ttwoecho~(\us)$                              & 12        & 7         &  5        & 12        & 14    & 8         & 28                & 12.3\\
Multiplexed average assignment fidelity, $F_{\text{RO}}~(\%)$       & 98.6      & 98.0      & 96.6      & 95.9      & 93.9  & 97.0      & 93.7              & 96.2\\
\hline
\end{tabularx}
\end{table*}

\section{Control electronics}
\begin{figure*}[ht]
\includegraphics[width = 0.85\textwidth]{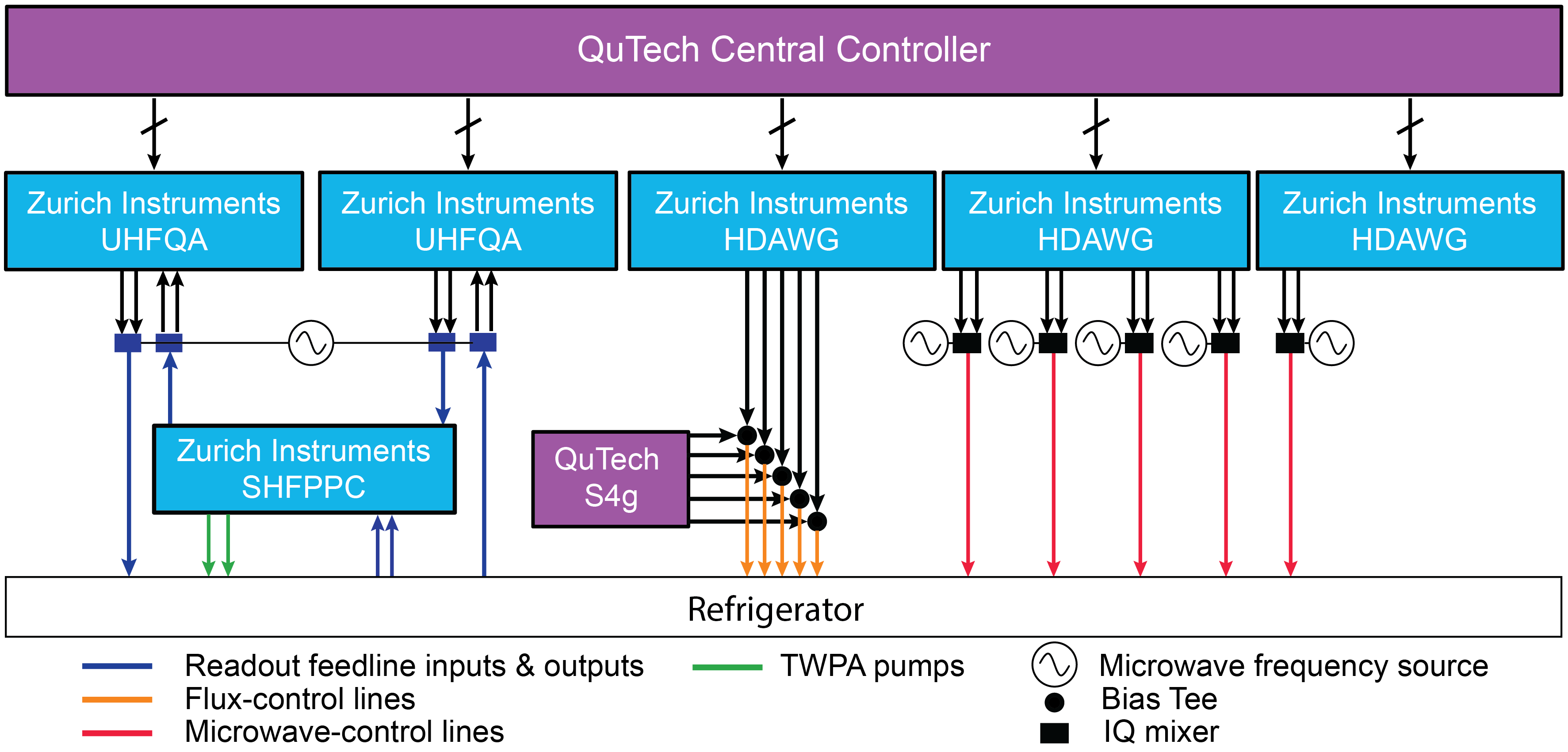}
\caption{
Schematic of the control electronics of $\Sfive$. For simplicity, small components such as attenuators (used to reduce reflection at mixers and bias tees) and directional couplers (used for mixer calibration) are not shown.
}
\label{fig:FigElectronics}
\end{figure*}

The room-temperature electronics of $\Sfive$ are shown schematically in Fig.~\ref{fig:FigElectronics}.
The electronics  of $\Sseven$ are identical, with only more channels used for flux biasing, flux pulsing and microwave driving.
QuTech S4G current sources provide the DC flux bias needed to tune all transmons to their simultaneous sweetspot.
Zurich Instruments (ZI) HDAWG8s generate the baseband flux pulses for two-qubit gates (not used here) and the envelopes for microwave qubit pulses.
ZI UHFQAs generate the envelopes of microwave pulses for readout, and digitize and process the response signals after down-conversion.
All pulse up- and down-conversions are performed using QuTech F1c2 IQ mixers, with Rohde$\&$Schwarz SGS100A microwave sources providing the local oscillators.
The above ZI analog front ends are triggered by the QuTech Central Controller (CC), an all-digital instrument timing all gate and readout operations on a $20~\ns$ grid.

TWPA pumps are generated by the ZI SHFPPC4, which also interferometrically cancels the transmitted pump tone at the input to the IQ down-conversion mixers.

\section{Data analysis and filtering}

\begin{figure}[ht]
\includegraphics[width=\columnwidth]{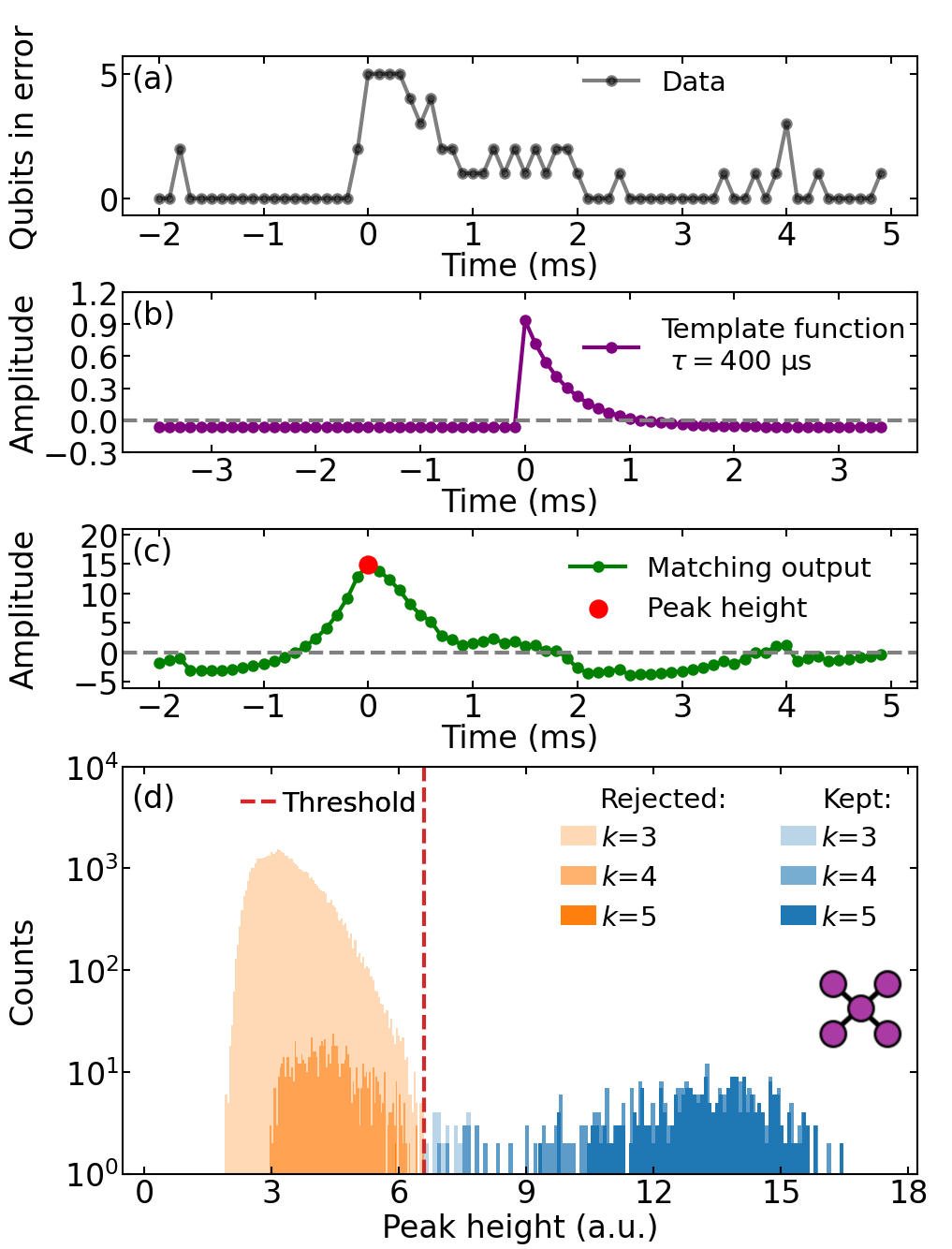}
\caption{
Illustration of the matched filtering method used for the categorization of all error events observed in $\Sfive$ and $\Sseven$.
(a) Example of an error event in $\Sfive$, with characteristic recovery time $\sim 2~\ms$ ($\tcycle=100~\us$ and $k=5$). (b,c)
The template matching of the error event with a (b) stepped exponential decay function of zero average gives the (c) matched-filter
output. We use the peak value of the matched-filter output to set $t=0$ for each error event. (d) Histogram of the matched-filter-output peak
value for all the error events in $\Sfive$ shown in Fig.~\ref{fig:Figure1}, with
$k=3,4$ and $5$. For each $k$, we observe two distinct distributions. The one with lower (higher) mean corresponds
to events to reject (keep). For each dataset, we set the filtering threshold to minimize false positives at $k=3$. We keep this threshold for all $k\geq3$.
}
\label{fig:cr_filtering}
\end{figure}

The binary-digitized measurement outcomes of all qubits are obtained using either of the circuit variants in Figs.~\ref{fig:Figure1}(a) or \ref{fig:Figure2}(a).
We define a single-qubit error as two back-to-back 0 outcomes, and an error event as $n$ qubits simultaneously in error, with $n\geq \nth$
[an example is shown for $\Sfive$ in Fig.~\ref{fig:cr_filtering}(a)]. The first step in the analysis identifies all error events, assigning them a timestamp.
The second step performs template matching on every identified event using a template function consisting of a stepped single-exponential decay (time constant $\tau$)
with an offset to null its average [Fig.~\ref{fig:cr_filtering}(b)], similar to~\cite{Wu25}. The peak output value of the template matching [Fig.~\ref{fig:cr_filtering}(c)] is used as the basis for keeping and rejecting events. Specifically, we use a histogram of all peak values to set the threshold that minimizes false positives for $\nth=3$, having earlier manually identified the $\tau$ that best separates the two apparent distributions.

\begin{figure}[ht]
\includegraphics[width=\columnwidth]{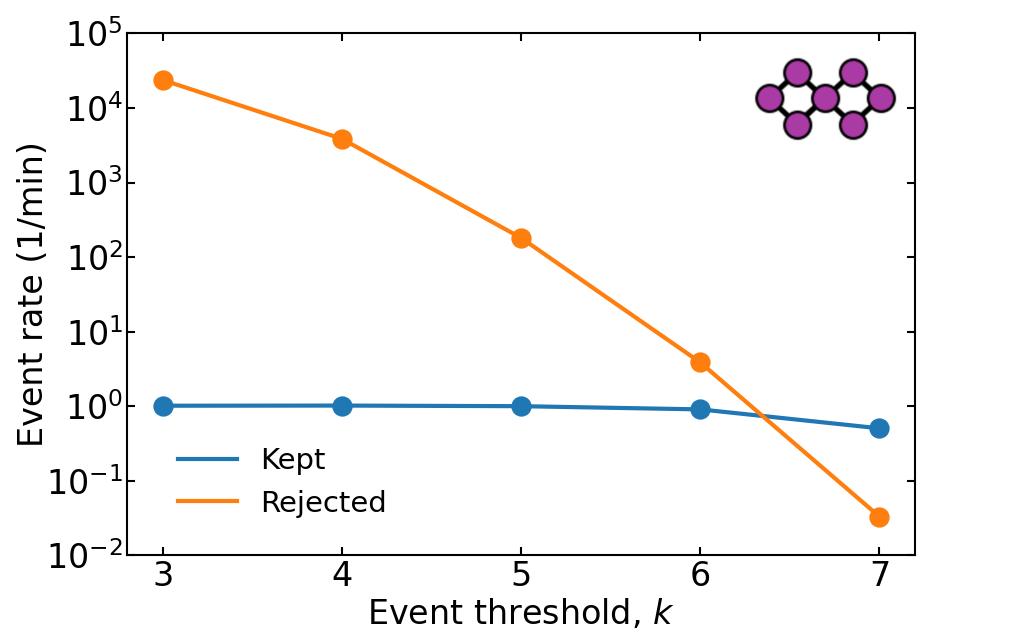}
\caption{
Comparison of the rates $\gammakept$ and $\gammarej$ of kept and rejected events, respectively, as a function of event threshold $\nth$,
for $\Sseven$ $(\tcycle=10~\us)$. The $\gammakept$ curve is the same as shown in Fig.~\ref{fig:Figure1}(e).
}
\label{fig:Figvsk}
\end{figure}

For completeness, Fig.~\ref{fig:Figvsk} shows the variation of the rate of rejected events $\gammarej$ with $k$ accompanying the variation in $\gammakept$ previously shown in Fig.~\ref{fig:Figure1}(e).

We also compare the $\gammakept$ of $\Sfive$ and $\Sseven$ to values reported by prior experiments using similar detection schemes (see~\cref{tab:TableRates}).
After accounting for device area, we observe that the $\Sfive$ rate is comparable to that in Ref.~\cite{Harrington24}, while the $\Sseven$ rate is comparable to those in Refs.~\cite{McEwen22, McEwen24}.

When studying the surges in $\Sfive$ (Figs.~\ref{fig:Figure4}, \ref{fig:FigSurgesA} and \ref{fig:FigSurgeFridgeB}), we perform an extra step of analysis to discern kept events with uncharacteristically long $\trec \gtrsim 2~\ms$.
We take the kept events obtained from the procedure above and perform a second round of template matching,
this time with $\tau =2~\ms$, using a new threshold to best discriminate such events from ones of typical duration.
(Additionally, all error events with an average baseline of simultaneous errors larger than $1.5$ are discarded.)

\section{Other experiments}

\subsection{Impact of cycle time on rates}
\begin{figure}[ht]
\includegraphics[width=\columnwidth]{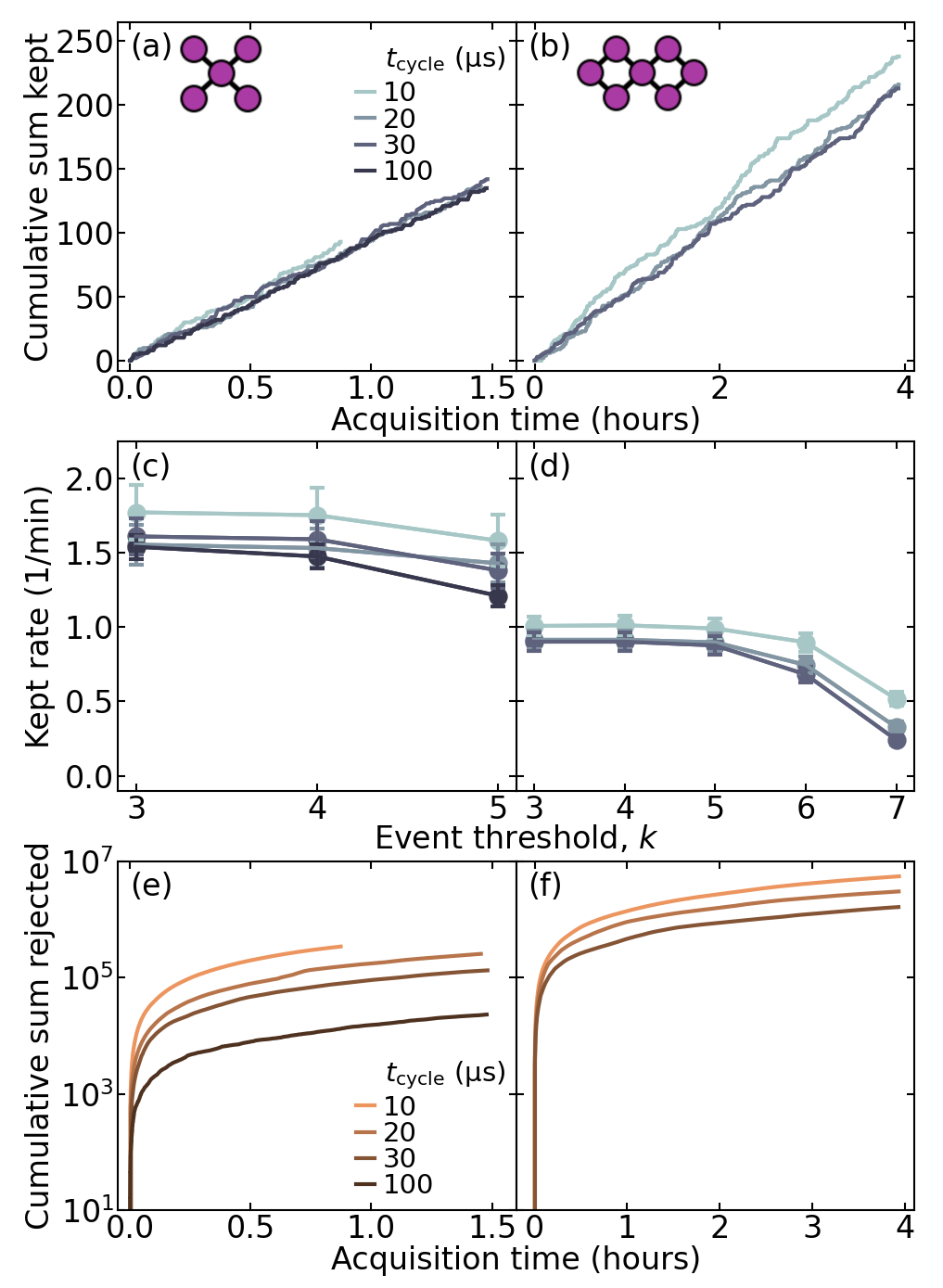}
\caption{
(a,b) Cumulative sums of kept events in (a) $\Sfive$ and (b) $\Sseven$ for $\tcycle=10$, $20$, and $30~\us$ in both devices, including $\tcycle=100~\us$ for $\Sfive$, and $\nth=3$.
(c,d) Extracted rate of kept events $\gammakept$ as a function of event threshold $k$ for (c) $\Sfive$ and (d) $\Sseven$.
(e,f) Corresponding cumulative sums of rejected events in (e) $\Sfive$ and (f) $\Sseven$. Note the logarithmic scale on the vertical axes.
All curves in this figure are obtained from the same raw data as in Figs.~\ref{fig:Figure2}(b,c).
}
\label{fig:vstcycle}
\end{figure}

Figure~\ref{fig:Figure2} investigated the effect of $\tcycle$ on the averaged response to error events, finding a dependence of $\trec$ only in $\Sfive$ which we attribute
to QP pumping. For completeness, Figure~\ref{fig:vstcycle} shows the corresponding impact on the rate of kept and rejected events, extracted from
the same raw data as in Figs.~\ref{fig:Figure2}(b,c) setting $\nth=3$. For $\Sfive$ and $\Sseven$, $\gammakept$ is independent of $\tcycle$ within the sensitivity of our measurements,
over the ranges covered ($10$ up to $100~\us$ for $\Sfive$, $10$ up to $30~\us$ for $\Sseven$). In both processors, $\gammarej$ decreases sharply with increasing $\tcycle$
[note semilog scales used in Figs.~\ref{fig:vstcycle}(e,f)]. This happens because (1) the probability per cycle of any qubit showing error from baseline $\Tone$ and
RO errors monotonically decreases with increasing $\tcycle$ and (2) the number of cycles per unit time evidently decreases as $1/\tcycle$.

\begin{table}[!ht]
\caption{
Comparison to error burst rates reported in the literature, normalized by chip area.
We restrict this comparison to experiments that use similar burst detection schemes.
}
\label{tab:TableRates}
\begin{tabularx}{\columnwidth}{p{3.4cm} p{2.1cm} p{2.7cm}}
\hline
Reference & Chip size & Normalized rate\\
 & ($\cm^{2}$) & $(\cm^{-2}~\minutes^{-1})$\\
\hline
This work ($\Sfive$)   & 0.64     & 2.41\\
This work ($\Sseven$) & 0.64      & 1.59\\
Harrington~\cite{Harrington24} & 0.25      & 2.23\\
McEwen (2024)~\cite{McEwen24} & 1      & 1.54\\
McEwen (2022)~\cite{McEwen22} & 4.8      & 1.25\\
\hline
\end{tabularx}
\end{table}

\subsection{Observed rates and QP pumping in refrigerator B}
\begin{figure}[ht]
\includegraphics[width=\columnwidth]{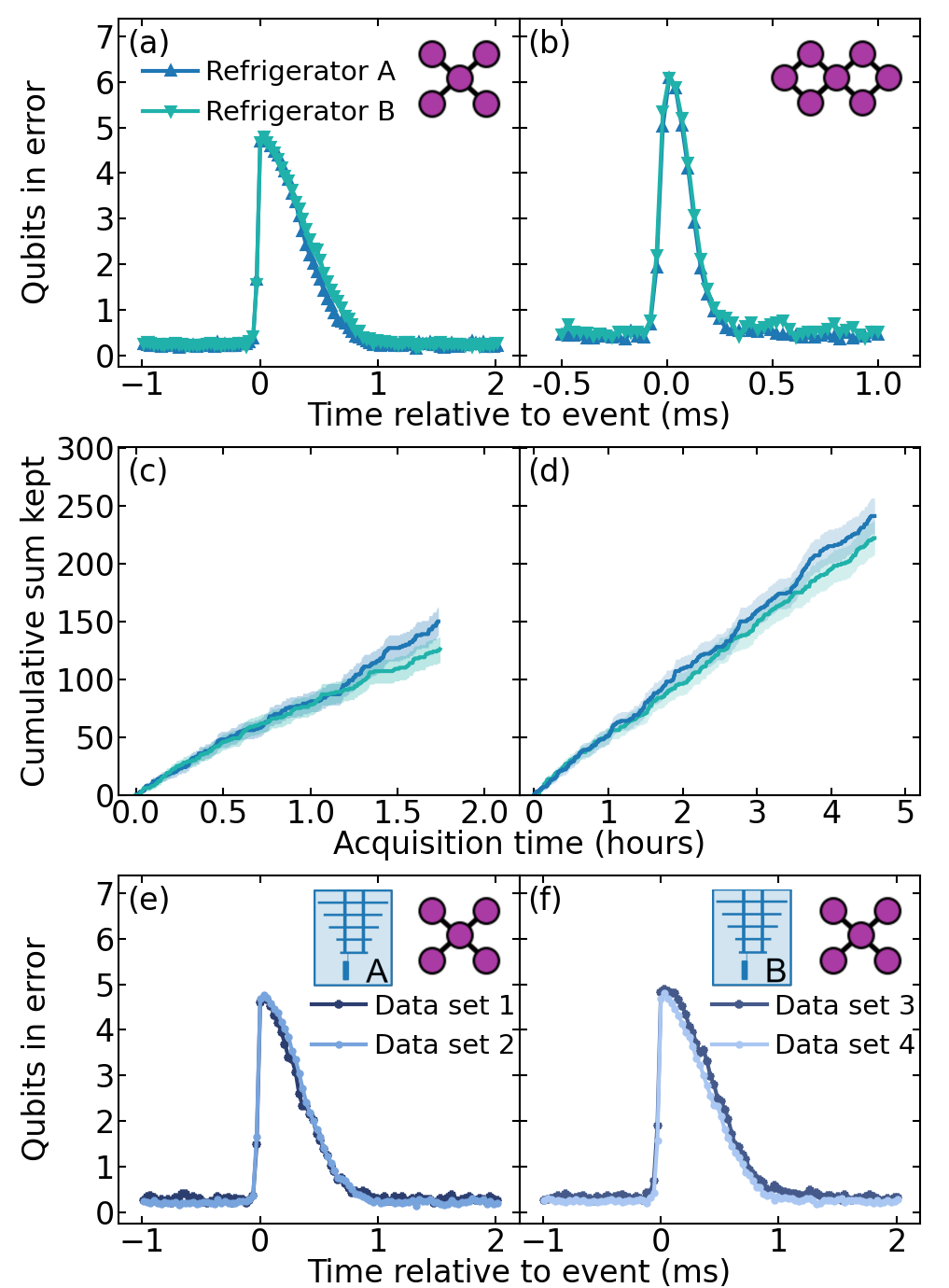}
\caption{
Comparison of kept events for each processor across refrigerators ($\tcycle = 30~\us$).
(a,b) Average of kept events for (a) $\Sfive$ $(\nth=5)$ and $\Sseven$ ($\nth=7$).
(c,d) Cumulative sum of kept events for (c) $\Sfive$ and (d) $\Sseven$ ($\nth=3$ for both).
The overlapping shaded areas represent the standard error of each cumulative-sum curve,
indicating that the rates in the two refrigerators are statistically indistinguishable.
(e,f) Average of kept events from two independent datasets for $\Sfive$ in (e) refrigerator A,
and in (f) refrigerator B showing variations within the same refrigerator.
These results show that $\trec$ and $\gammakept$ are refrigerator independent.
In (c), the curve for refrigerator A is the same one shown for shield off in Fig.~\ref{fig:FigLead}(a).
}
\label{fig:FigKeptFridges}
\end{figure}

\begin{figure}[ht]
\includegraphics[width=\columnwidth]{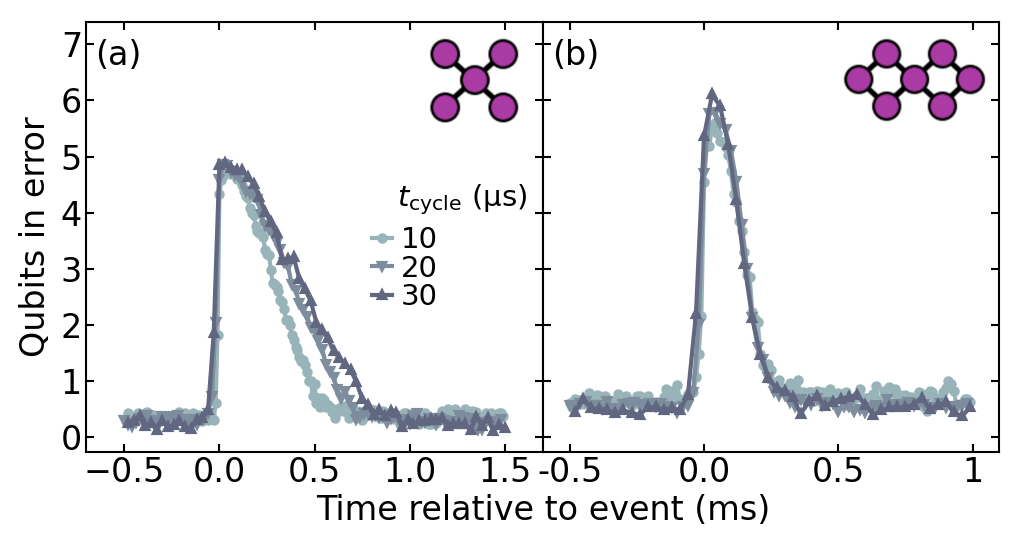}
\caption{
We also performed the experiment of Fig.~\ref{fig:Figure2}(a) with the devices in refrigerator B.
(a,b) Average of kept events for (a) $\Sfive$ and (b) $\Sseven$.
We observe that $\tcycle$ affects $\trec$ only in $\Sfive$, with the same signature as observed in refrigerator A.
}
\label{fig:FigPumpingSwap}
\end{figure}

To test whether the main observations show any refrigerator dependence, we performed similar experiments before and after moving $\Sfive$ and $\Sseven$
from refrigerator A to refrigerator B, not simultaneously but in succession.
The control electronics always followed the processor. We find that $\gammakept$, $\trec$ and the QP pumping
effect are refrigerator independent, as shown conclusively in Figs.~\ref{fig:FigKeptFridges} and \ref{fig:FigPumpingSwap}.
Apparent differences in the average of kept events in Fig.~\ref{fig:FigKeptFridges}(a) were also
observed in independent datasets obtained within the same refrigerator, as shown in Fig.~\ref{fig:FigKeptFridges}(e,f).
We conclude that these variations are not refrigerator-specific.

\subsection{Impact of a rudimentary lead shield}
\begin{figure}[ht]
\centering
\includegraphics[width=\columnwidth]{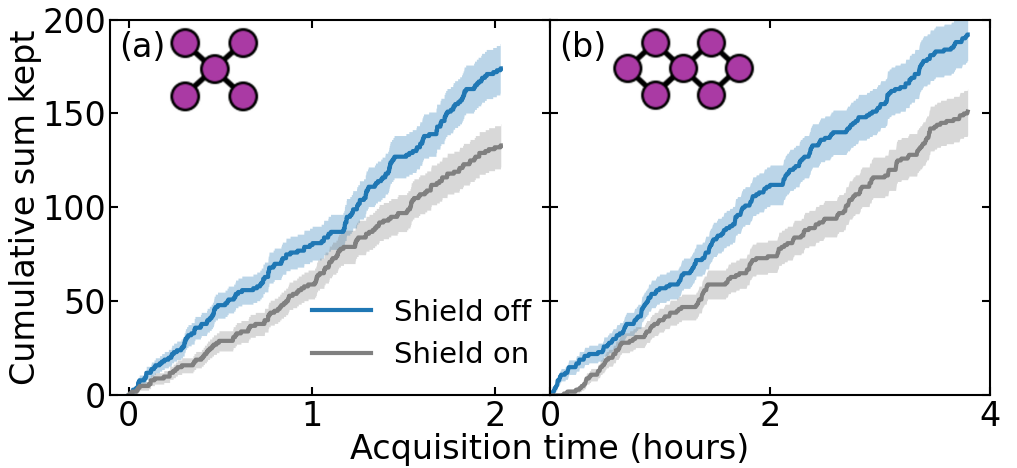}
\caption{
Cumulative sum of kept events for (a) $\Sfive$ (b) $\Sseven$ with lead shield on and off in refrigerator A ($\tcycle=30~\us$ and $\nth=3$).
The shield reduces $\gammakept$ by $(24\pm9)\%$ [$(21\pm9)\%$] in $\Sfive~[\Sseven]$.
In panel (a), the shield-off curve is the same one shown for in Fig.~\ref{fig:FigKeptFridges}(c).
}
\label{fig:FigLead}
\end{figure}

Inspired by prior experiments~\cite{Vepsalainen20, Li24}, we repeat the experiment of Fig.~\ref{fig:Figure1} using a
rudimentary Pb shield to affect $\gammakept$. Our shield consists of multiple sheets ($\sim 3.5~\mm$ thickness each) that we wrap around the outer vacuum chamber (OVC) of refrigerator A, amounting to a total of $\sim 1~\cm$ thickness. Figure~\ref{fig:FigLead} compares $\gammakept$ for both processors in two configurations: shield on (as described above) and shield off
(meaning shield lowered clear of the OVC).  The shield-on configuration reduces $\gammakept$ by $(24\pm9)\%$ in $\Sfive$, and $(21\pm9)\%$ in $\Sseven$.
Previous work~\cite{Li24} using a rectangular $1~\cm$-thick lead shield reported a $(32\pm5)\%$ rate reduction, larger than observed here, and most likely the result of better shield coverage.

\subsection{Other example observations of the surge in $\Sfive$ in refrigerator A}
\begin{figure*}[!ht]
\includegraphics[width=\textwidth]{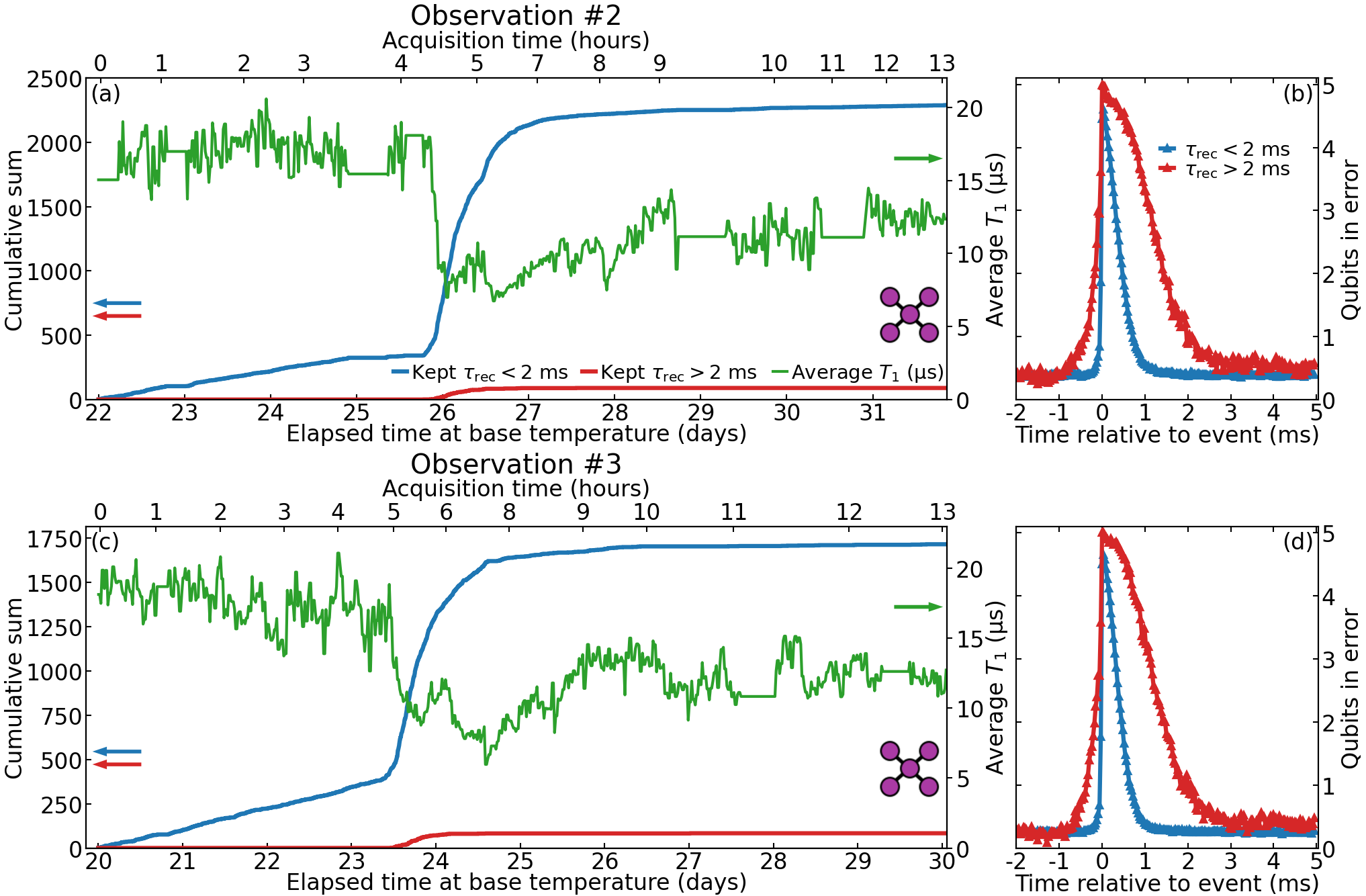}
\caption{
Other observations of the surge in $\Sfive$ with $\tcycle = 30~\us$ while in refrigerator A. (a,b) Observation $\#2$. (c,d) Observation $\#3$. These surges show  similar characteristics to the surge shown in Fig.~\ref{fig:Figure4}, which was observation $\#4$.
}
\label{fig:FigSurgesA}
\end{figure*}

We have observed the surge in $\Sfive$ in refrigerator A a total of 4 times prior to moving the device to refrigerator B.
The first time (observation $\#1$, data not shown), we became aware of the surge only once it was underway.
Since then, we have observed the effect following every thermal cycle. Observations $\#2$ and $\#3$ are presented
in Fig.~\ref{fig:FigSurgesA}, with similar features as observation $\#4$ shown in Fig.~\ref{fig:Figure4}. We list some details on observations $\#2$, $\#3$ and $\#4$ below.

$Observation~$\#2$~-~$ Following a thermal cycle to $4~\K$, the refrigerator reached base temperature on 24/06/2024.
After calibrating readout and single-qubit gates, we gradually and sequentially calibrated the two-qubit gates
over  $22$~days. The latter calibrations took $\sim 2~\hours$ per qubit pair and involve repeated flux pulsing of the transmons~\cite{Negirneac21}.
By spacing them out, we could observe that the repeated flux pulsing on all transmon pairs was not the trigger of the surge.
The surge began on 18/07/2024.

$Observation~$\#3$~-~$ Following a thermal cycle to $4~\K$, the refrigerator reached base temperature on 15/08/2024.
Once again, we calibrated readout and single-qubit gates first, and then sequentially calibrated two-qubit
gates over $17$~days. The surge began only on 08/09/2024.

$Observation~$\#4$~-~$ Following a thermal cycle to $4~\K$, the refrigerator reached base temperature on 02/10/2024.
We calibrated readout, single-qubit gates, and the two-qubit
gates sequentially over $6$~days. The surge began on 28/10/2024.

All these observations follow a similar pattern. Initially, $\gammakept$ is steady
(e.g., Fig.~\ref{fig:Figure1}) for $3-4$ weeks upon reaching base temperature.
When the surge begins, $\gammakept$ sharply increases by a factor $\sim 10$ for $\sim 12~\hours$, and then monotonically decreases.
Within a few days, it stalls to a value $\sim 100$ times lower than the initial rate. In each case, the stall rate persisted until the device was thermal cycled. We find that thermal cycle to $4~\K$ fully resets the effect.
Interestingly, the Al JJ electrodes (and also airbridges, crossovers and wirebonds) become normal
conducting, but all circuit elements defined by the NbTiN base layer (including the floating transmon
capacitor pads) remain superconducting.

In retrospect, the surge must have been the reason why, upon first learning of catastrophic simultaneous-error
events measured by the Google team~\cite{McEwen22}, we were unable to immediately observe the effect in $\Sfive$.
At the time, $\Sfive$ was publicly available through the Quantum Inspire platform and had been at base
temperature and online for several months. In all likelihood, the surge had already happened and $\gammakept$ vanished.

\subsection{Example observation of the surge in $\Sfive$ in refrigerator B}
\begin{figure*}[ht]
\includegraphics[width=\textwidth]{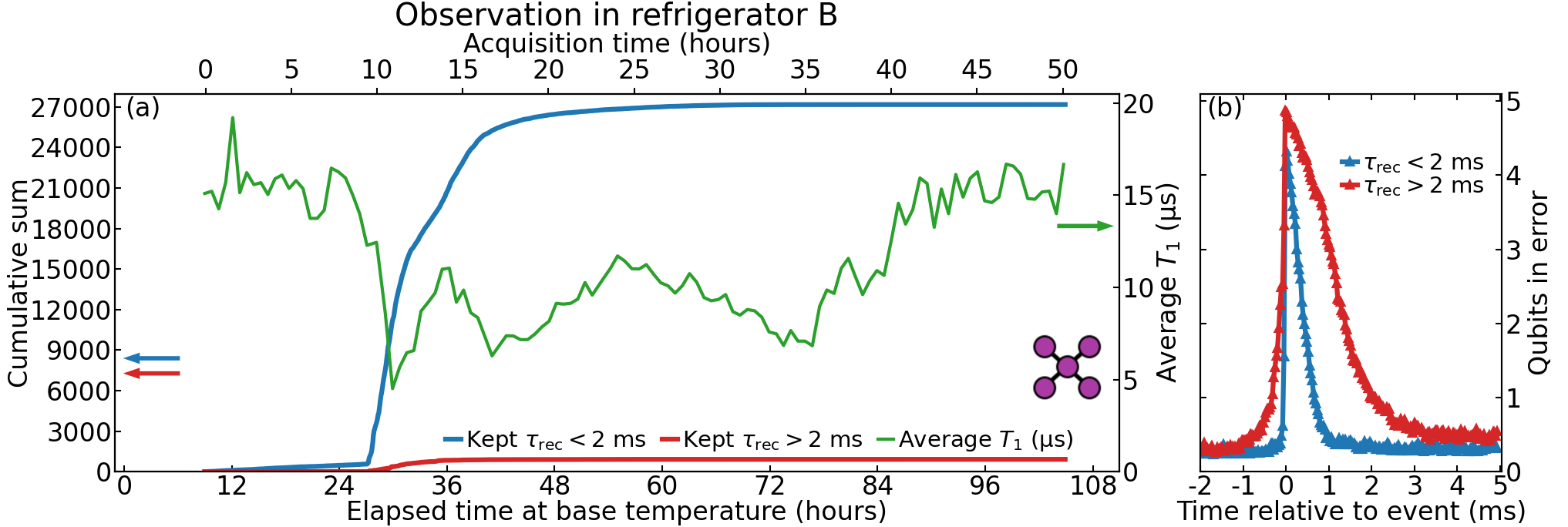}
\caption{
Example observation of the surge in $\Sfive$ with $\tcycle = 30~\us$ in refrigerator B. While the surge has similar characteristics to those observed in refrigerator A
(Figs.~\ref{fig:Figure4} and \ref{fig:FigSurgesA}), there are notable differences. In this observation (and in 11 of 12 observations in refrigerator B to date),
the surge begins just $\sim 1$~day after reaching base temperature ($\sim1~$week for the $12^{\mathrm{th}}$). At the start of the surge, $\gammakept$ increases by a factor $\sim 100$, compared to  $\sim 10$ in refrigerator A. This feature is common to the 12 observations in refrigerator B.
}
\label{fig:FigSurgeFridgeB}
\end{figure*}

The surge in $\Sfive$ also occurs in refrigerator B, however much less reliably and predictably than in refrigerator A.
In fact, we have observed it in $12$ out of $20$ cooldowns. In $11$ of cases, the surge started just $\sim 1$~day after reaching base temperature, while in the remaining case it started after $~\sim1$~week. Cooldowns during which the surge did not occur maintained base temperature over $4$ to $42$~days. As shown by the example observation in Fig.~\ref{fig:FigSurgeFridgeB}, the onset of the surge in refrigerator B shows an increase of $\gammakept$ by a factor $\sim 100$ compared to just $\sim10$ in refrigerator A.

\end{bibunit}

\end{document}